\documentclass[11pt,oneside,reqno]{amsart}
\usepackage{comment}
\usepackage{amscd}
\usepackage{amssymb}
\usepackage{amsthm}
\usepackage{amsmath}
\usepackage{latexsym}
\usepackage{float}
\usepackage{empheq,mathtools}
\usepackage{amstext}
\usepackage{graphicx}
\usepackage{color}
\usepackage{hyperref}
\usepackage{a4wide}

\usepackage[english]{babel}

\usepackage[pagewise]{lineno}

\usepackage{tabularx,booktabs,caption}
\newcolumntype{C}[1]{>{\Centering}m{#1}}

\theoremstyle{plain}

\theoremstyle{definition}

\title[Modeling and Simulation of COVID-19]{Modeling and Simulation of the spread of coronavirus disease (COVID-19) in Lebanon}

\author[Ayman Mourad]{Ayman Mourad$^1$}
\address[Ayman Mourad]{\newline
	  Department of Mathematics, Faculty of Sciences (I), Lebanese University, Hadat, Lebanon}
\email[]{ayman.mourad@ul.edu.lb}

\author[Fatima Mroue]{Fatima Mroue$^2$}
\address[Fatima Mroue]{\newline
	Department of Mathematics, American University of Beirut, Beirut, Lebanon}
\email[]{fm47@aub.edu.lb}

\date{
\today\\
$^1$Department of Mathematics, Faculty of Sciences (I), Lebanese University, Hadat, Lebanon\\
$^2$Department of Mathematics, American University of Beirut, Beirut, Lebanon
}
\begin{document}
\begin{abstract}
In this paper, we develop a probabilistic mathematical model for the spread of coronavirus disease (COVID-19). It takes into account the known special characteristics of this disease such as the existence of infectious undetected cases and the different social and infectiousness conditions of infected people. In particular, it considers the social structure and governmental measures in a country, the fraction of detected cases over the real total infected cases, and the influx of undetected infected people from outside the borders. Although the model is simple and allows a reasonable identification of its parameters, using the data provided by local authorities on this pandemic, it is also complex enough to capture the most important effects. We study the particular case of Lebanon and use its reported data to estimate the model parameters, which can be of interest for estimating the spread of COVID-19 in other countries. We show a good agreement between the reported data and the estimations given by our model. We also simulate several scenarios that help policy makers in deciding how to loosen different measures without risking a severe wave of COVID-19. We are also able to identify the main factors that lead to specific scenarios which helps in a better understanding of the spread of the virus.
\end{abstract}

%\subjclass{}

%\keywords{}
\maketitle
%%%%%%%%%%%%%%%%%%%%%%%%%%%%%%%%%%%%%%%%%%%%%%%%%%%%%%%%%%%%%%%%%%%%%%%%%%%%%%%
\section{Introduction}
After the identification of a novel strain of coronavirus in China in December 2019, later labeled COVID-19, the virus spread in China and most of the countries of the world. The World Health Organization (WHO) declared the outbreak as a public health emergency of international concern \cite{world2005statement} by January 2020 and the situation was declared as a pandemic by March 2020\cite{whoPandemic}. The global spread of the virus has lead to an unprecedented infectious disease crisis worldwide with 15,666,671 confirmed cases and 636,787 deaths by July 30, 2019.\\
With the uncertainties about the COVID-19 virus, governments must put in place measures to minimize mortality rates as well as the economic impact of the viral spread \cite{anderson2020will}.  In this context, predictive mathematical models for epidemics are fundamental, for policy makers to take the right decisions in order to minimize mortality and morbidity and flatten the epidemic curve for the sake of reducing the strains on the healthcare system,  in a timely way taking into account the socioeconomic consequences \cite{anderson2020will}. Moreover, the effects of mitigation strategies significantly vary depending on population characteristics such as age structure, the stage of the epidemic such as the number of infected and immunized people and the level of population compliance with the measures. Therefore, model-based predictions of the eventual effect of different mitigation measures on the epidemic, taking into consideration such local parameters, are crucial to support evidence-based decisions.\\
Various models have been proposed in the literature for disease spreads. They can be categorized into agent-based models (ABM) \cite{bonabeau2002agent,ajelli2010comparing} and compartmental models \cite{hethcote2000mathematics,diekmann2012mathematical,brauer2012mathematical}. Compartmental models are built on differential equations and assume that the population is perfectly mixed with people moving between compartments such as susceptible (S), infected (I) and recovered (R)\cite{kermack1927contribution,naasell1996quasi,hurley2006basic,jin2007sirs}. These models revealed the threshold nature of epidemics and triumphed in explaining ``herd immunity". However, they fail to capture complex social networks and the behavior of individuals who may adapt depending on disease prevalence. On the other hand, agent-based models can capture irrational behavior and complex social networks and are used to simulate the interactions of autonomous agents that can be either individuals or collective entities \cite{epstein2009modelling}. Furthermore, ABM approaches in epidemiology can simulate such complex dynamic systems with less simplification of the rich variation among individuals. Also, the statistical variance is more evident than in compartmental models, whose smooth curves often misleadingly express more certainty than justified, due to the randomization at each run \cite{hunter2017taxonomy,hunter2018open,tracy2018agent,kai2020universal}.\\
Several models have been developed for the COVID-19 pandemic. Some are compartmental models such as \cite{lin2020conceptual,anastassopoulou2020data,casella2020can,giordano2020modelling} while others are stochastic transmission models \cite{hellewell2020feasibility,kucharski2020early} and agent-based models \cite{chang2020modelling,kai2020universal}. In the present work, we propose a stochastic individual agent-based model for the spread of COVID-19. The model takes into account the characteristics of the virus and attributes to them probability distributions. We mention, for instance, the incubation period, infectiousness, testing sensitivity, ... It also incorporates probability distributions for social and individual conditions such as: family size within and outside the same household, number of people contacted per day and their subdivision into known versus unknown contacts... These distributions are fitted for the case of Lebanon but they can be modified depending on the country. In Section \ref{sec:description}, we describe the parameters used in the model and the rational of using them. Section \ref{sec:simulation} contains the calibration of the model for the case of Lebanon, the validation of the results obtained by the simulations by comparing to the real data and the predictions of the strength of the epidemic by the end of August and September depending on the mitigation measures imposed by policy makers such as the closure or opening of the airport and the strictness of measures in the country. Finally, we conclude with some remarks in Section \ref{sec:conclusion}.

%%%%%%%%%%%%%%%%%%%%%%%%%%%%%%%%%%%%%%%%%%%%%%%%%%%%%%%%%%%%%%%%%%%%%%%%%%%%%%%
\section{Description of The Model}\label{sec:description}
\begin{figure}[h!]
  \centering
  \includegraphics[width=14cm]{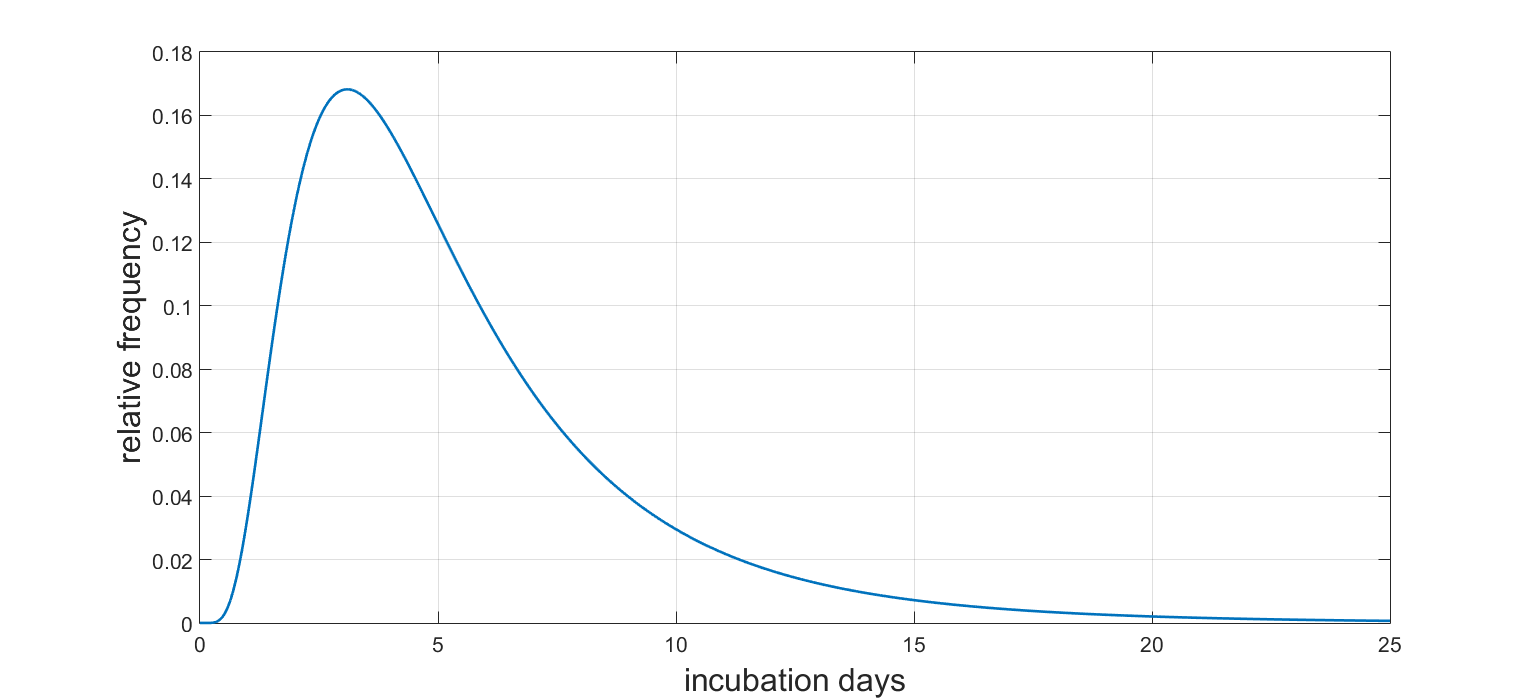}  
   \caption{PDF for the incubation period (lognormal distribution) }\label{Incubation}
\end{figure}

In this model, we consider each individual as an agent with variety of variables that are considered relevant to model COVID-19 spreading. In particular, to each infected individual are attributed the following parameters: 
\begin{itemize}
\item[-]the times of infection and detection: it is important to reduce the delay between the infection time and the detection time because during this period, infected individuals remain in the community while possibly infecting others \cite{klinkenberg2006effectiveness}.\\

\item[-]the probability of infection (transmitting the virus to others depending whether the person is wearing a mask, obeying hygiene measures, keeping a safe distance, ...): this probability differs also depending on the nature of contact between the infected person and his contacts where we differentiate family members in the same household from contacts outside the house such as at work or in public transportation. The probability of infecting family members sharing the same house is assumed to be higher during the period extending from the instant of infection to the time of detection.\\

\item[-]the incubation period (time from exposure to illness onset): the general incubation period for COVID-19 varies for different individuals and ranges from 2 to 14 days. It is 5-6 days on average with the 95$^{th}$ percentile of the distribution at 12.5 days and the log-normal distribution provides the best fit to the data for incubation period estimates  \cite{li2020early,linton2020incubation}(see Figure \ref{Incubation}).\\

%\item[-]the serial interval (duration from onset of symptoms in an infector (a primary-case patient) to the onset of symptoms in an infectee (a secondary-case patient)): the mean serial interval for COVID-19 was estimated as 7.5 days in \cite{li2020early} but more recent studies \cite{nishiura2020serial,du2020serial} suggest that it is around 4 days. Being shorter than the mean incubation period, pre-symptomatic transmission is likely and may be more frequent than symptomatic transmission \cite{nishiura2020serial,giordano2020modelling}. We use as suggested in \cite{nishiura2020serial,du2020serial} a normal distribution for the serial interval estimates (see Figure \ref{Serial}).\\

\item[-]the connections: those are either family members or contacts of the individual such as coworkers, neighbors,... To account for family members we consider a log-normal probability distribution function for the average family size or household size (see Figure \ref{Family} for the case of Lebanon). While for other contacts, we consider an exponential distribution for the daily new encounters. So if a person was in contact with the individual on the first day, he is not counted in the new encounters of the following days (see Figures).\\

\item[-]contact-tracing: the identification of the source of infection, tracing and isolating its contacts are crucial for breaking the chain of transmission and for the control of the epidemic \cite{hellewell2020feasibility}. Two parameters influence the effectiveness of contact-tracing and isolation: the transmissibility of the pathogen (measured by the basic reproductive number) and the proportion of presymptomatic transmission \cite{fraser2004factors}. Very high levels of contact tracing are required in the case of presymptomatic infectiousness \cite{hellewell2020feasibility}. In the present model, we account for contact tracing efficiency by including a probability distribution for the proportion of newly infected people whom source of infection is unknown. We also consider the probability of a detected infected person not to restrict to isolation measure and keep roaming around irresponsibly.\\

\item[-]the localization: in order to account for the effect of airport and border opening, we split individuals into travelers and locals. For travelers, we distinguish between those having a positive screening result and those with a negative result. Those with a positive test result are assumed to be quarantined and are no longer infectious. For the negative result group, we consider the probability of having a false negative test result in addition to the percentage of those who do not go into a period of isolation. On the other hand, the local infected residents are subdivided into two groups: symptomatic with positive test result versus those screened due to their contact with an infected person before onset of symptoms. The probability distribution function for travelers is based on data collected from local health authorities.\\ 
\end{itemize} 
We note that the ABM approach  needs  realistic  data, typically obtained from a census, to create agents that effectively capture human behavior \cite{epstein2007controlling}. 

%\begin{figure}[h!]
%  \centering
%\includegraphics[width=14cm]{serial_normal.png}
%\caption{PDF for the serial interval (normal distribution)}\label{Serial}
%\end{figure}
%\begin{figure}[h!]
%  \centering
%\includegraphics[width=14cm]{household_pdf_lognormal.png}
%\caption{PDF for the family size (log-normal distribution)}\label{Family}
%\end{figure}
%%%%%%%%%%%%%%%%%%%%%%%%%%%%%%%%%%%%%%%%%%%%%%%%%%%%%%%%%%%%%%%%%%%%%%%%%%%%%%%
\section{Numerical Simulations and Calibration of The Model for The Case of Lebanon}\label{sec:simulation}
\subsection{Overview of the situation in Lebanon}
The first confirmed case of COVID-19 in Lebanon was detected on February 21, 2020 and by February 29, there was a total of 7 confirmed cases going back from travel trips. Consequently, educational institutions were closed, and gradual measures were introduced until the declaration of general mobilization and state of emergency by mid-March where public gatherings were banned, cultural venues closed and social distancing measures imposed in public. Furthermore, the airport, the land borders and the seaports were closed as of March 19. On April 5, new arrivals were intermittently allowed through the airport. By the end of May, authorities in Lebanon have been gradually easing restrictions. Public transportation has resumed, with social-distancing measures. Government institutions and certain private companies, including various shops and stores, were permitted to return to normal operations from June 1 and the lockdown was lifted on June 7. Starting July 1, flights were resumed and  the airport started operating up to 10 $\%$ of its capacity \cite{ministry}. Travelers with negative PCR results upon arrival were isolated for a maximum of 3 days. However, people were not abiding by the preventive measures, and travelers were not respecting the isolation period\cite{arabnews}. The virus rebounded to reach a total number of 4205 by July 29 and the authorities reinstated lock-down from July 30 to August 3 and from August 6 to August 10. In this context, it is obvious that there is an urgent need for evidence-based decisions particularly that the country is facing an unprecedented economic crisis.
\subsection{Calibration}
\begin{figure}[h!]
  \centering
\includegraphics[width=14cm]{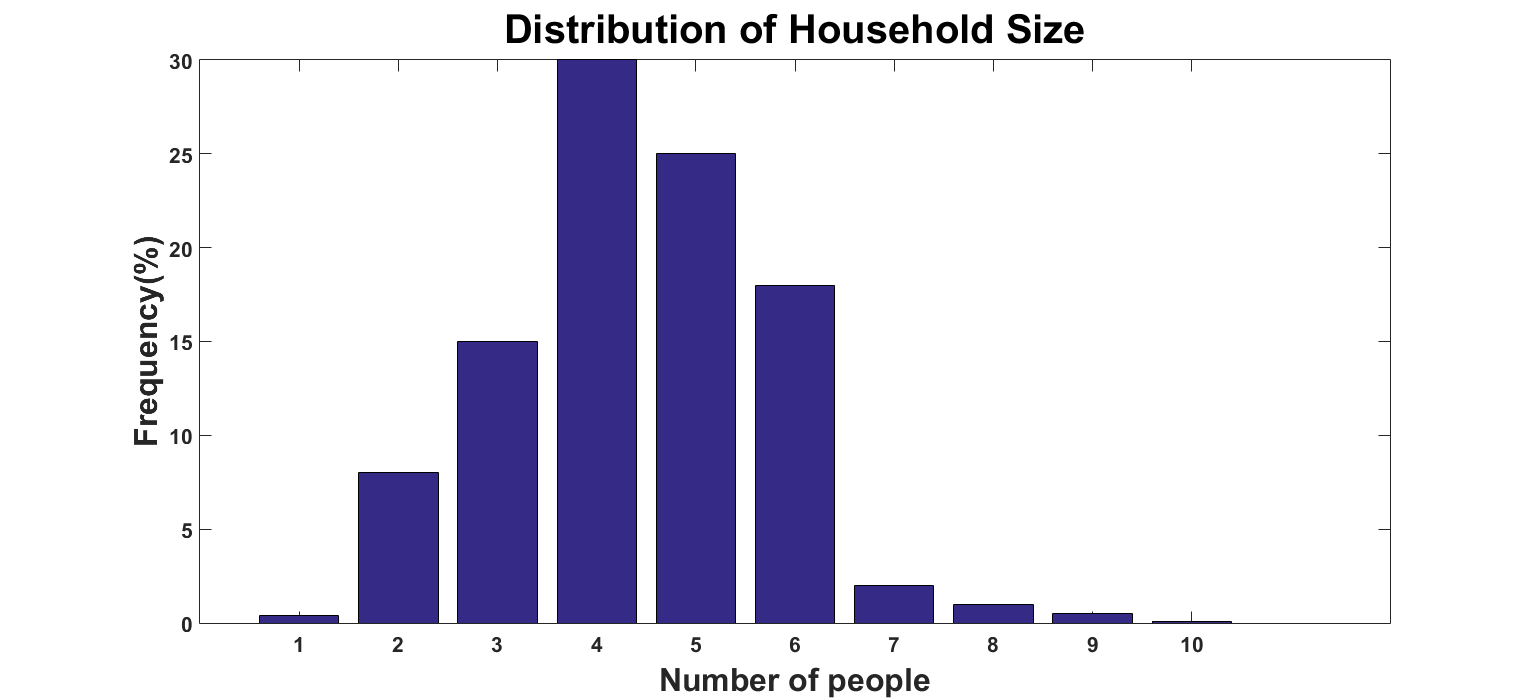}
\caption{Distribution for the family size in the same household.}\label{Family}
\end{figure}

In order to obtain a realistic model for the case of Lebanon, first we used part of the real data to estimate the parameters that are involved in the model. Then we validated the model for the second part of the real data. Indeed, using numerical optimization techniques, we calibrated the model and computed the parameters in order to reproduce results that are close to real data for the period before June 30, published on the website of the ministry of public health of Lebanon (https://coronanews-lb.com/). During this period, strict measures regarding the travelers were imposed in Lebanon, and the number of social events (such as weddings) was still small. On the other hand, as of July 1$^{st}$,  measures were relaxed and people were less respectful to the measures so the average number of new people met per day was increased, the quarantine period for travelers was restricted to 3 days and the percentage of travelers respecting the quarantine duration decreased.\\
For the incubation period, we used a discrete distribution obtained from \cite{li2020early,linton2020incubation} as in Figure \ref{Incubation}. On the other hand, due to the lack of census in Lebanon, we used the reports \cite{masri2008development,yaacoub2012population} to deduce a distribution for family size in the same household (see Figure \ref{Family}). In the absence of official reports, the distribution of new people met per day was also deduced from our familiarity with the Lebanese context where public transport is not prevalent, the number of family members met per day is high, ... \\

\subsection{Validation}
\begin{figure}[h!]
  \centering
\includegraphics[width=14cm]{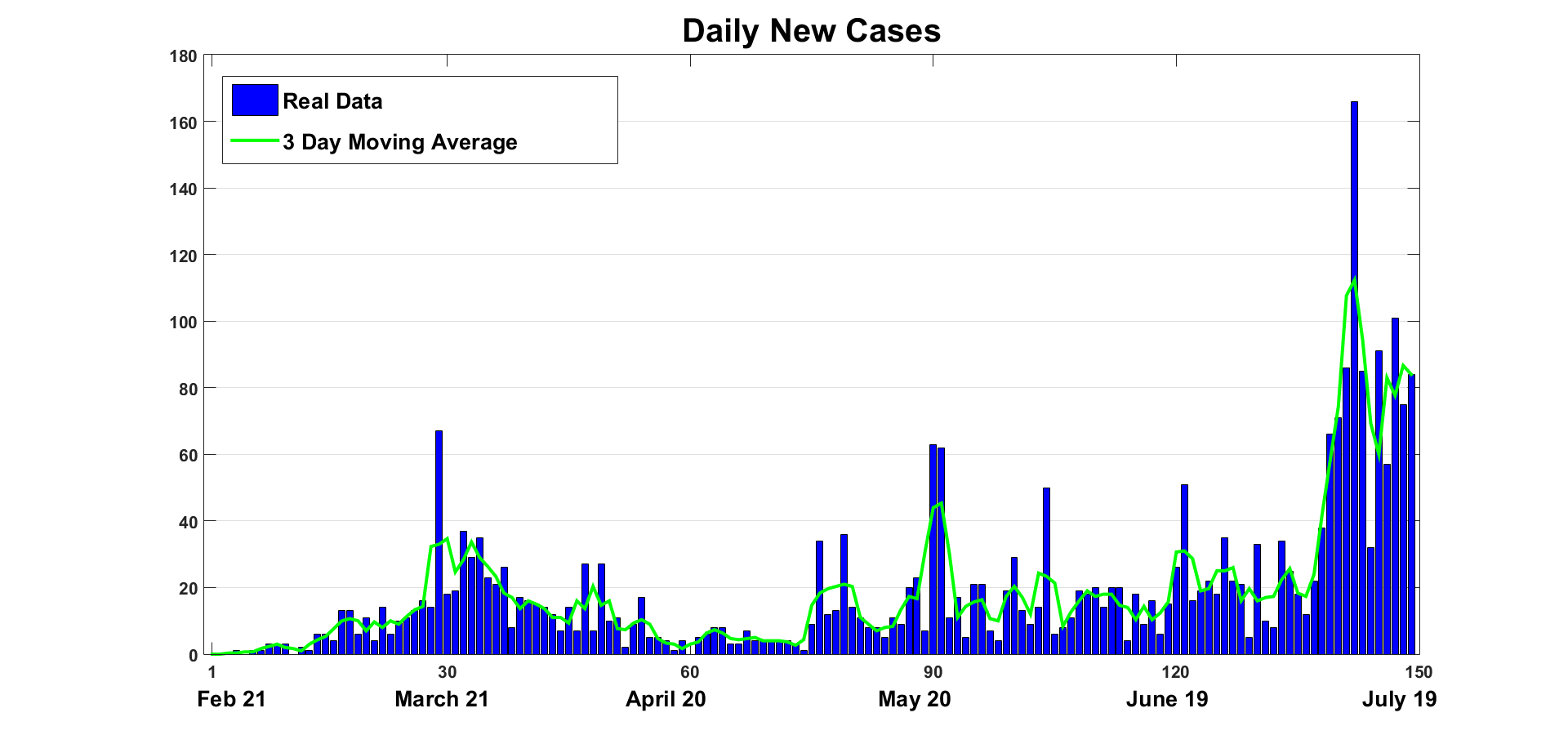}
\caption{Daily new cases as reported by the ministry of public health in Lebanon from February 21 till July 19.}\label{DailyNewReal}
\end{figure}

\begin{figure}[h!]
  \centering
\includegraphics[width=14cm]{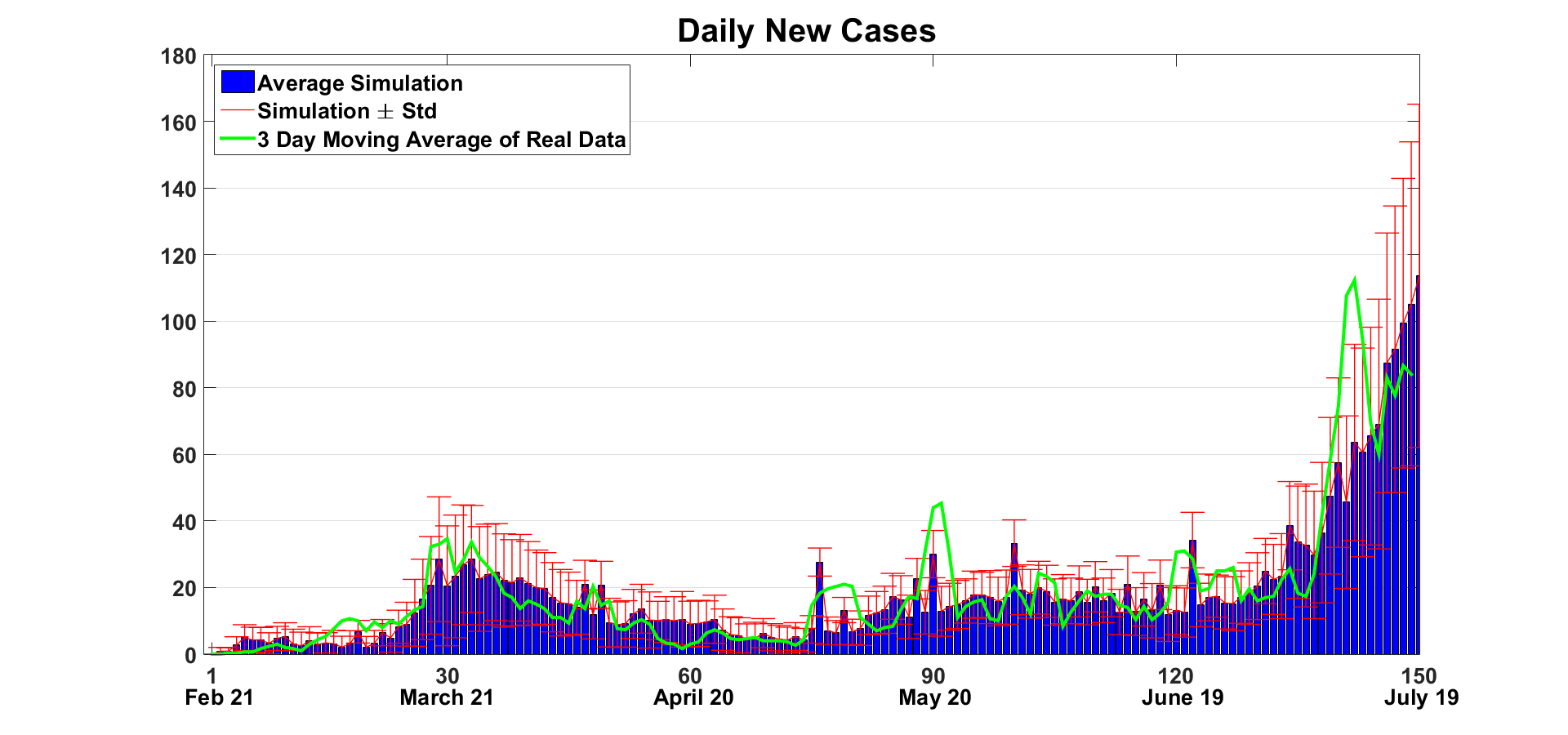}
\caption{Daily new cases from February 21 till July 19 as obtained by the simulation with the red bars showing the obtained values $\pm$ the standard deviation.}\label{DailyNewSimJuly19}
\end{figure}

\begin{figure}[h!]
  \centering
\includegraphics[width=14cm]{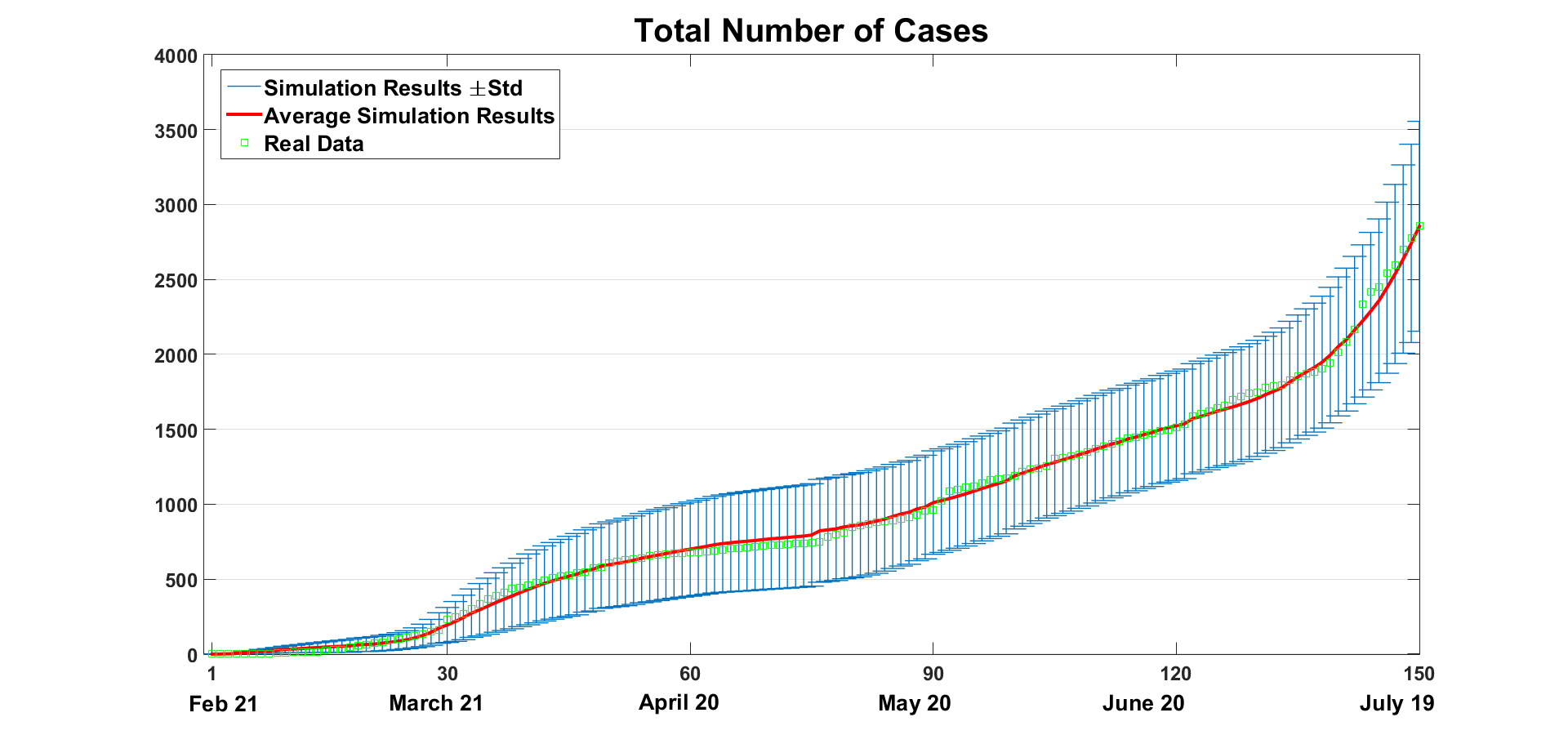}
\caption{Cumulative number of cases as reported by the ministry of public health in Lebanon from February 21 till July 19 (in green) and as obtained by the simulation (in red).}\label{CumulJuly19MeanStd}
\end{figure}

The simulation results that we obtained for the suggested model are very close to the real data for the period up to June 30 due to the parameters estimation. Moreover, these results are close to the real data for the period from July  1$^{st}$ to July 19 which validates the model.\\
Figure \ref{DailyNewReal} shows the daily new confirmed cases as reported on the website of the ministry of public health in Lebanon. Figure \ref{DailyNewSimJuly19} shows the daily new cases from February 21 till July 19 obtained by the simulation with the aforementioned parameters and Figure \ref{CumulJuly19MeanStd} shows the cumulative number of cases per day for both the real data and the simulations.\\
In order to obtain a representative probabilistic result, the simulations were obtained from 60 runs of the probabilistic model, so we represent the result as the average number of daily new cases within the range of the corresponding standard deviation.


\subsection{Prediction}
The prediction of the time course of the epidemic being crucial for policymakers. We simulate in this paragraph four scenarios that differ according to the measures (strict or loose) and to the airport (open or closed) for the periods of the first two weeks and the last two weeks of August.\\

\begin{figure}[h!]
  \centering
\includegraphics[width=7cm]{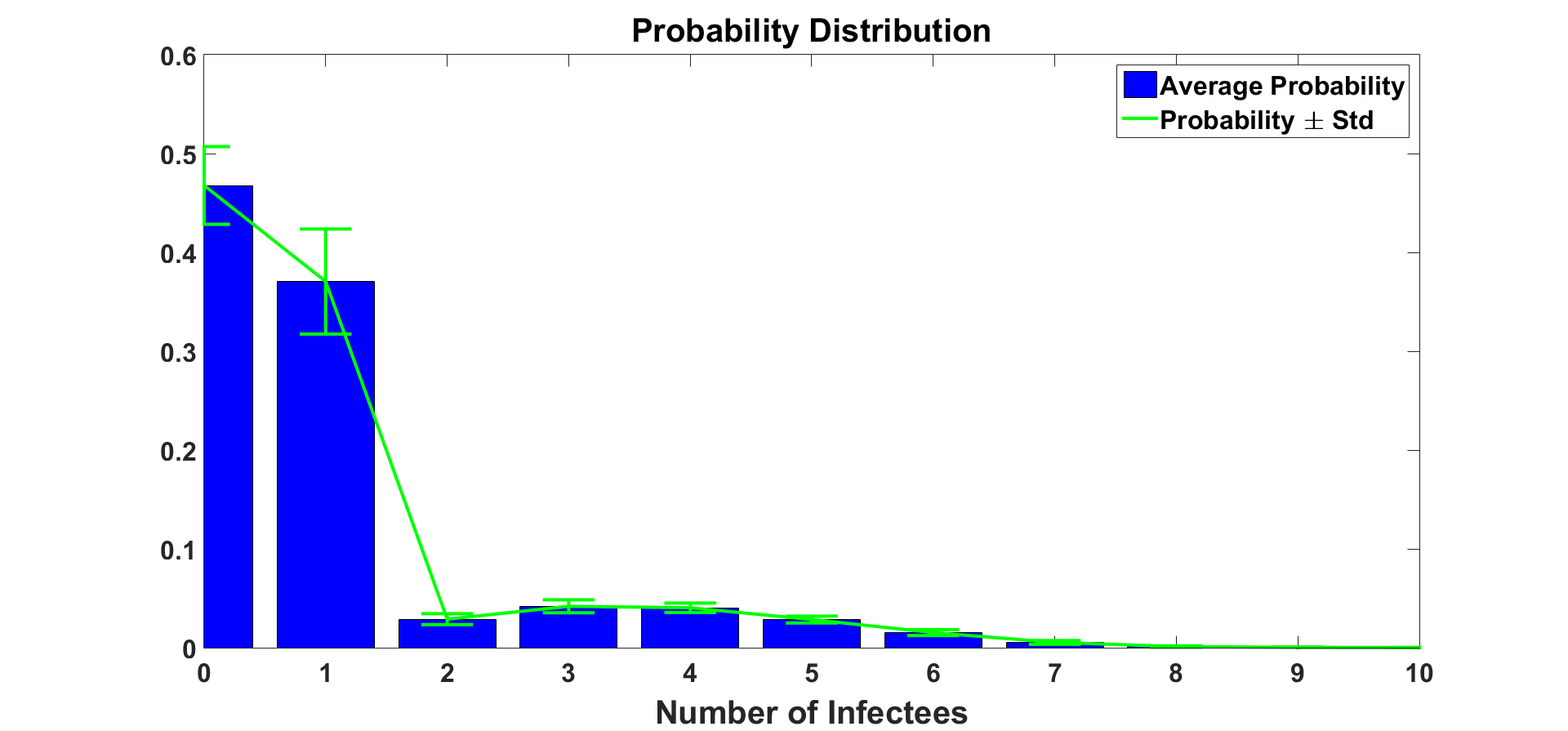}
\hspace*{0.2cm}
\includegraphics[width=7cm]{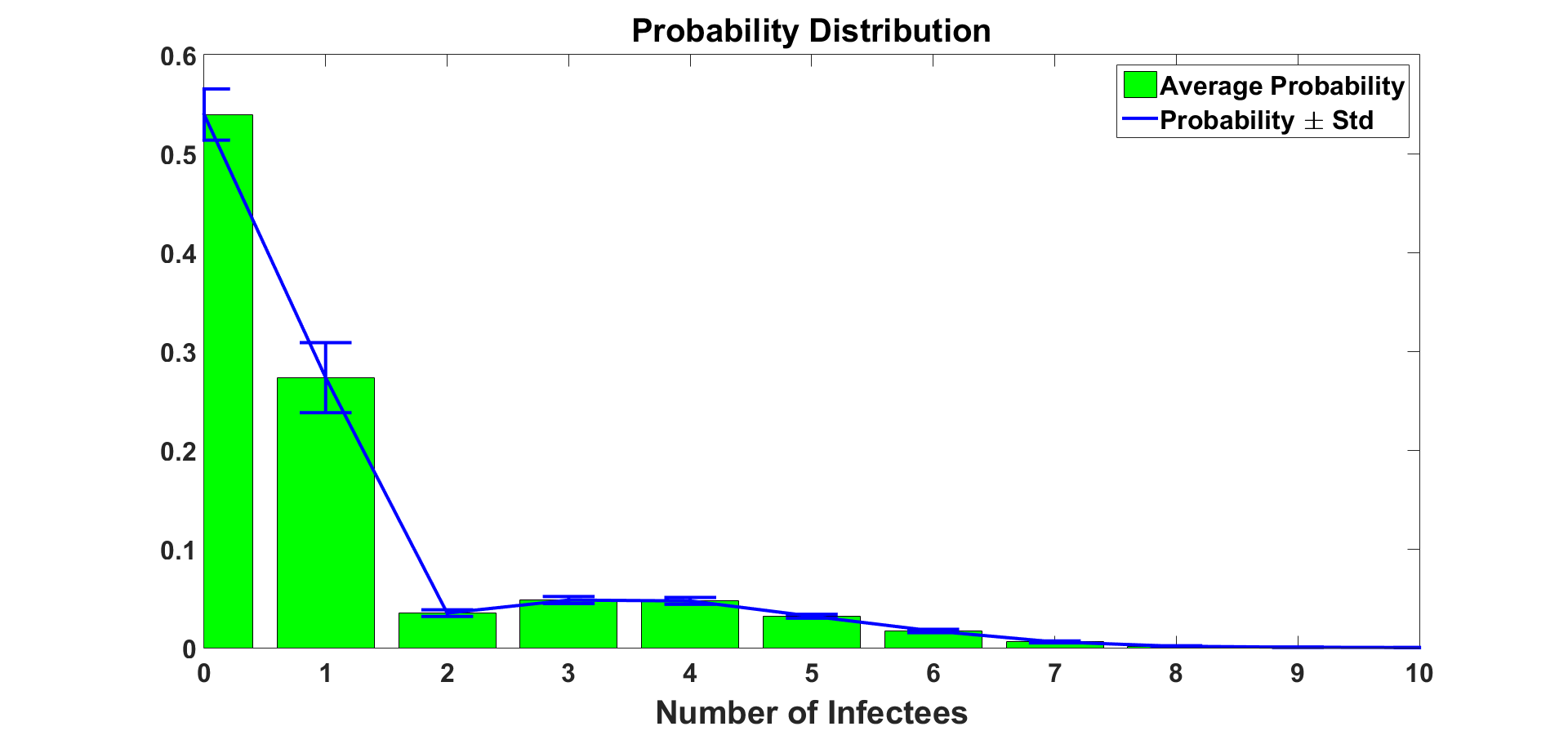}
\caption{Probability distribution for the number of people that an infected person infects before July 1$^{st}$(left) and between July 1$^{st}$ and August 31 (right), obtained from the model provided that strict measures are imposed and the airport is kept open during August.}\label{Infection per Person osos}
\end{figure}

\begin{figure}[h!]
  \centering
\includegraphics[width=14cm]{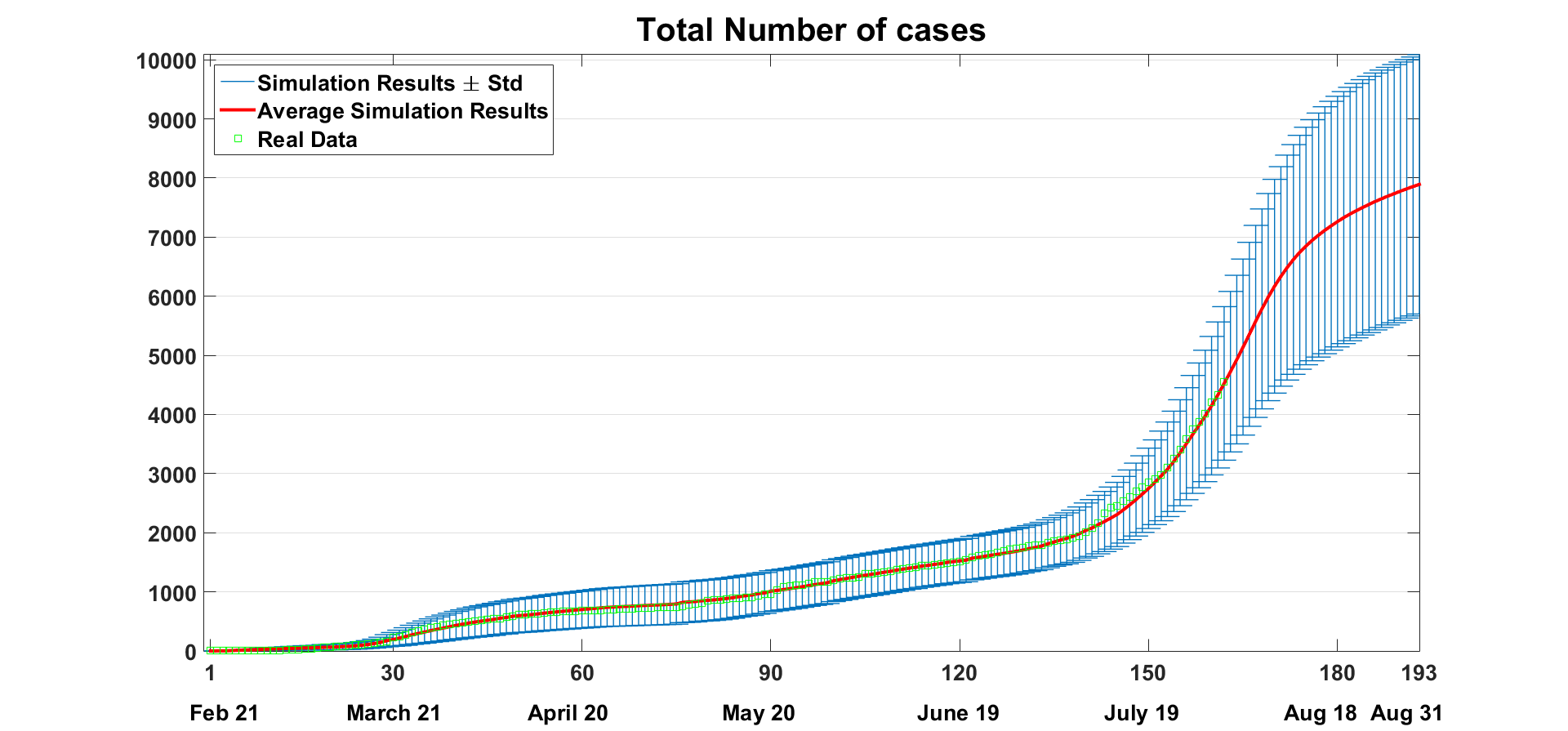}
\caption{Scenario I: Strict measures and airport open during August. Total number of cases.}\label{ososCumul}
\end{figure}

\begin{figure}[h!]
  \centering
\includegraphics[width=14cm]{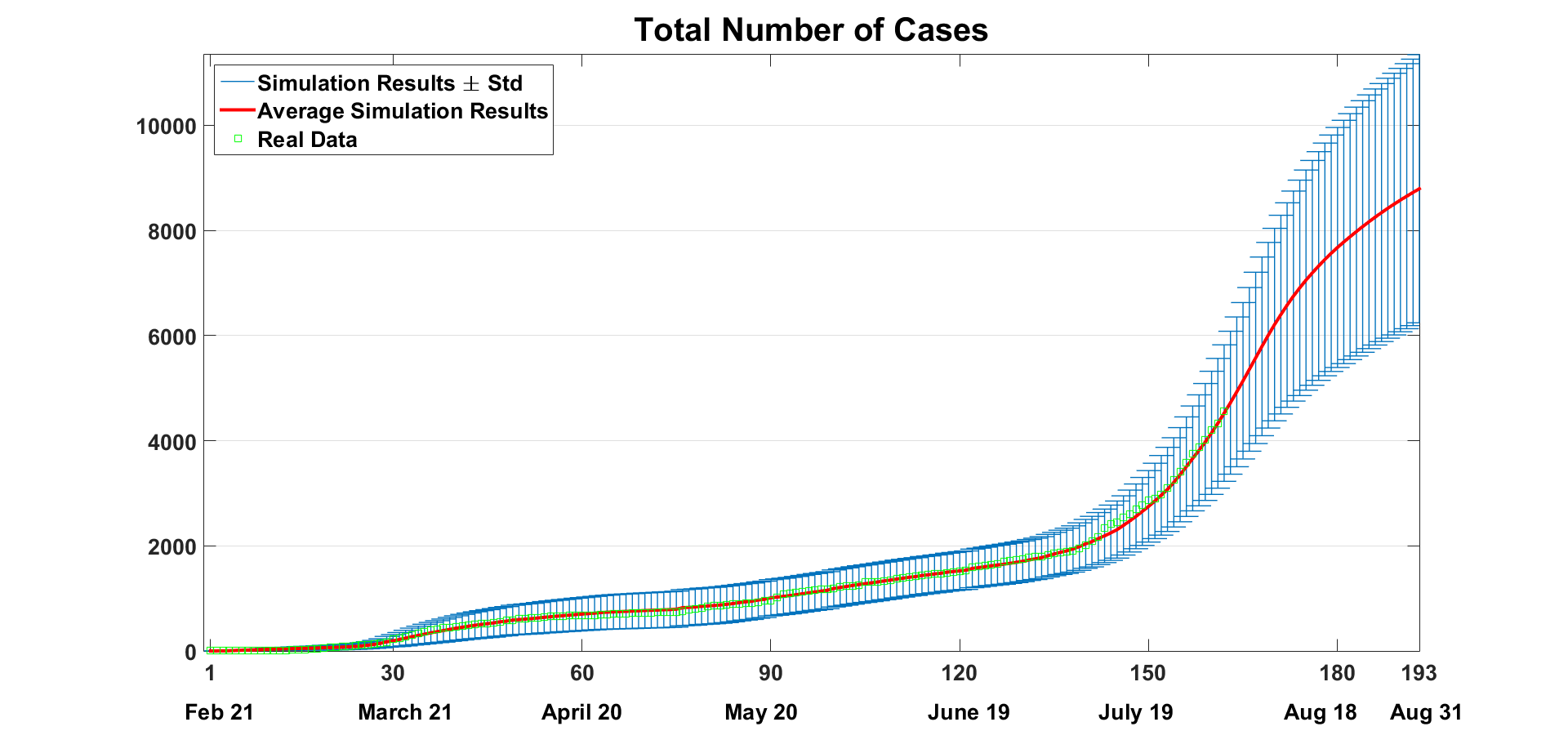}
\caption{Scenario II: No strict measures and airport open during August. Total number of cases.}\label{ononCumul}
\end{figure}

First, we consider that during August, strict measures are adopted by citizens but the airport kept open. This scenario (Scenario I) corresponds to the government policy in response to the bound in the number of cases during the last week of July. Figure \ref{ososCumul} shows the curve of the total number of cases until the end of August. This number reaches by the end of August 7,893 cases on average $\pm$ 2,193. Figure \ref{Infection per Person osos} shows the effectiveness of strict measures in decreasing the number of people infected by the same person as shown in the higher probability of a person not to infect any other person when strict measures are imposed during August.\\
Second, we consider that during August, less stringent measures are adopted by citizens and they are mostly behaving as in the period following the lift of the lock-down. Moreover, we assume that the airport is kept open. Figure \ref{ononCumul} shows the result of the simulation of this scenario (Scenario II) where the total number of cases reaches by the end of August 8,793 cases on average $\pm$2,549.\\

We also consider two more scenarios: Scenarios III and IV where strict measures and no strict measures are imposed during all of August respectively, but for both scenarios the airport closes for the last two weeks of August.

\begin{figure}[h!]
  \centering
\includegraphics[width=7cm]{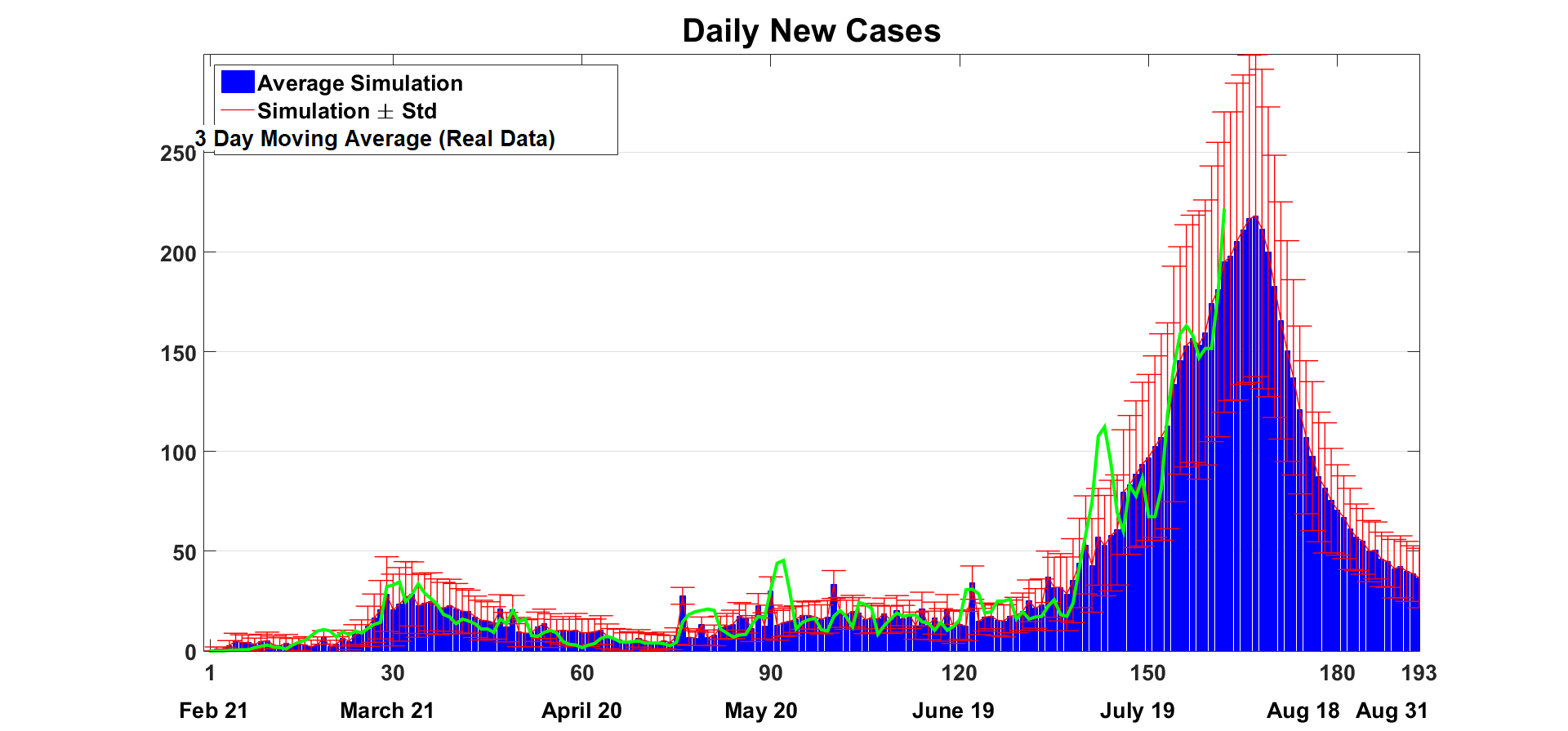}\hspace*{.2cm}
\includegraphics[width=7cm]{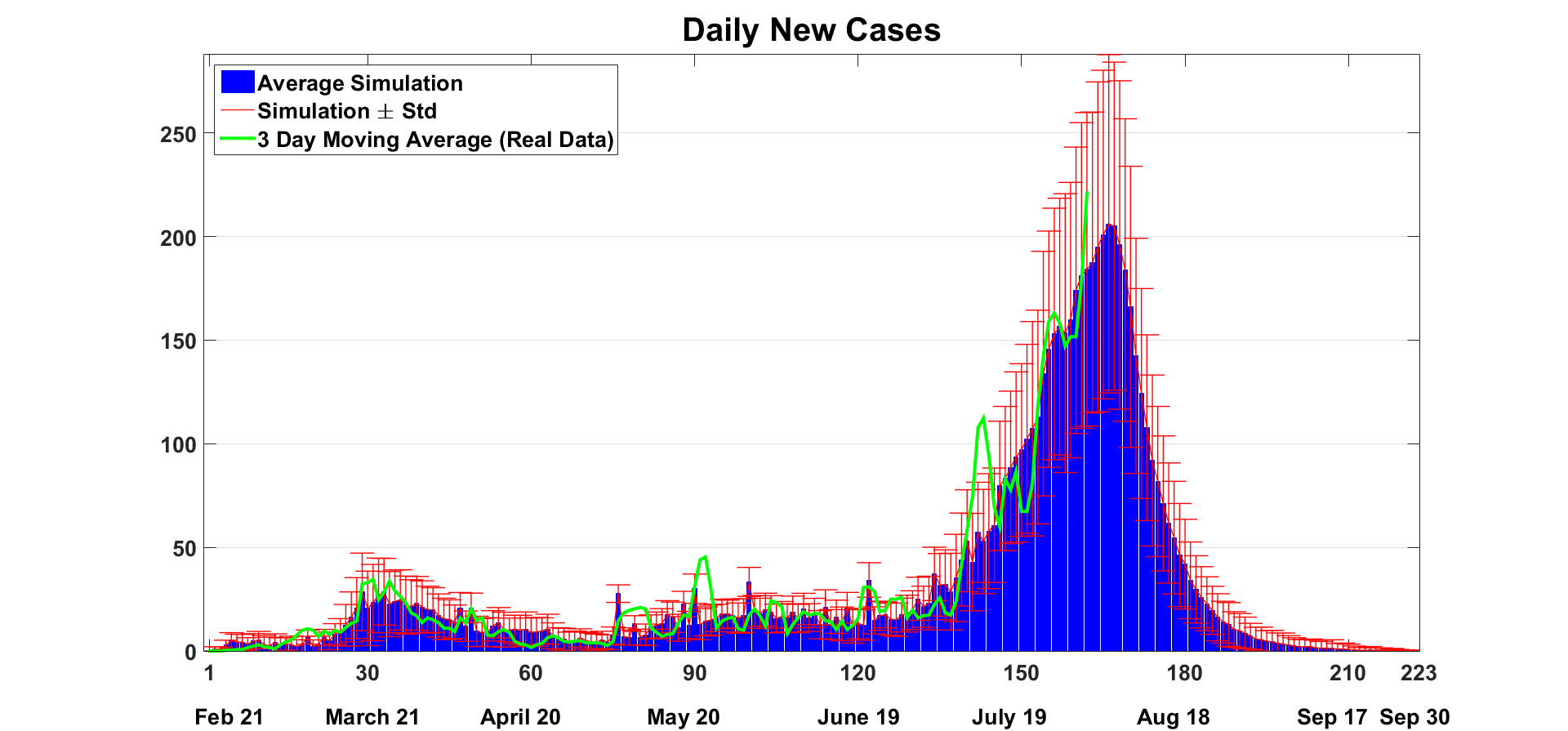}\\
\includegraphics[width=7cm]{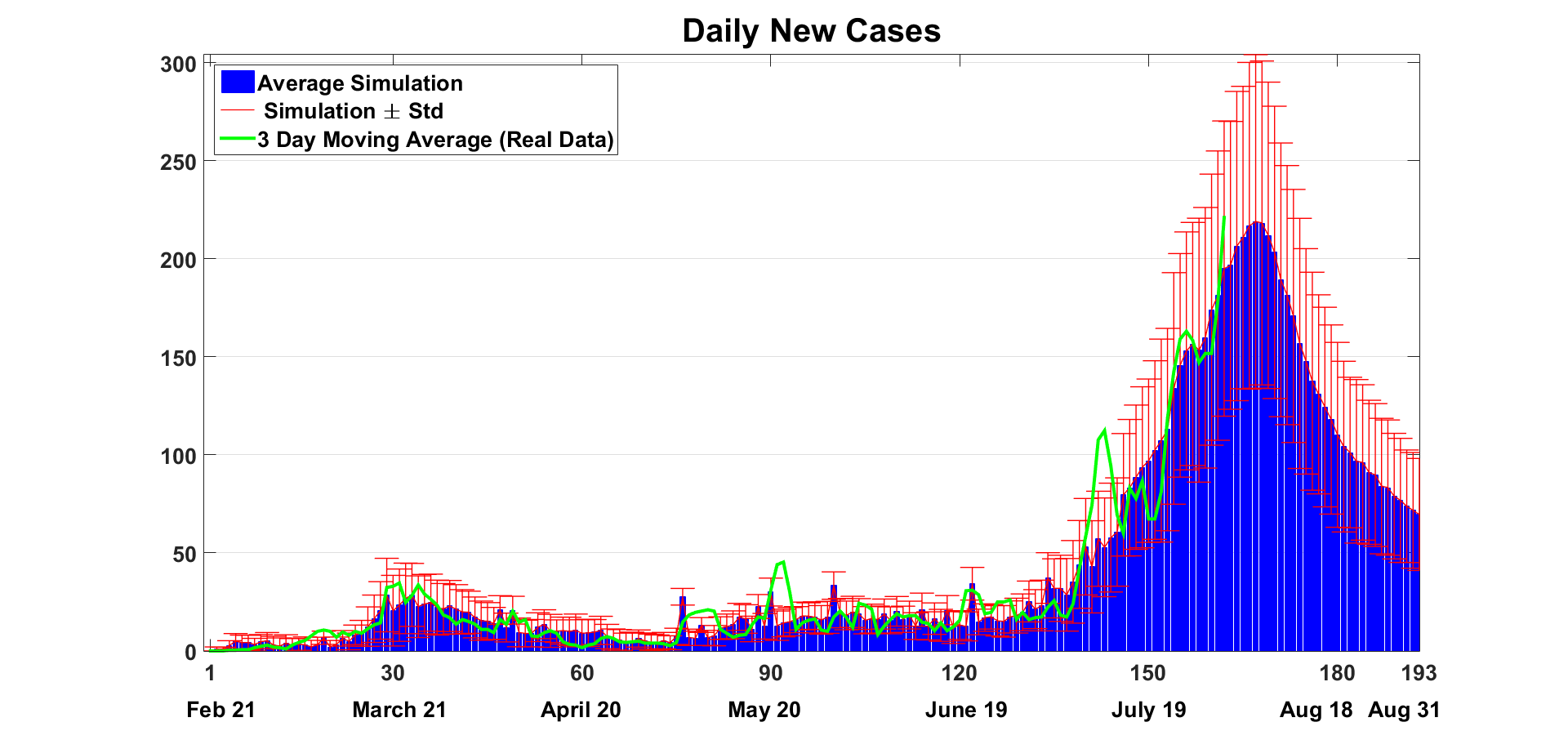}\hspace*{.2cm}
\includegraphics[width=7cm]{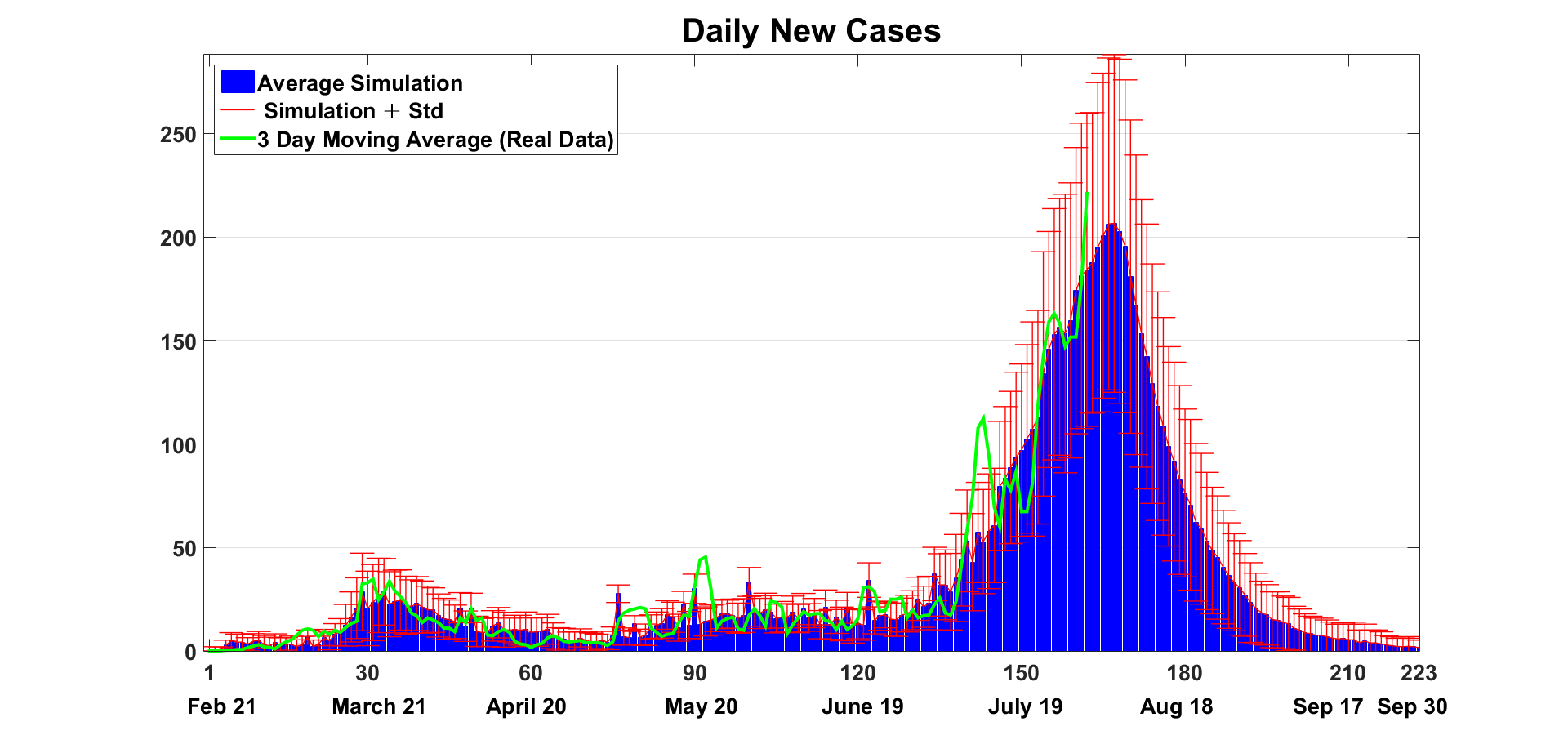}
\caption{Simulation results: Daily new cases for the four scenarios. Top-left: Scenario I, bottom-left: Scenario II, top-right: Scenario III, bottom-right: Scenario IV.}\label{Daily No Measures}
\end{figure}

\begin{figure}[h!]
  \centering
\includegraphics[width=7cm]{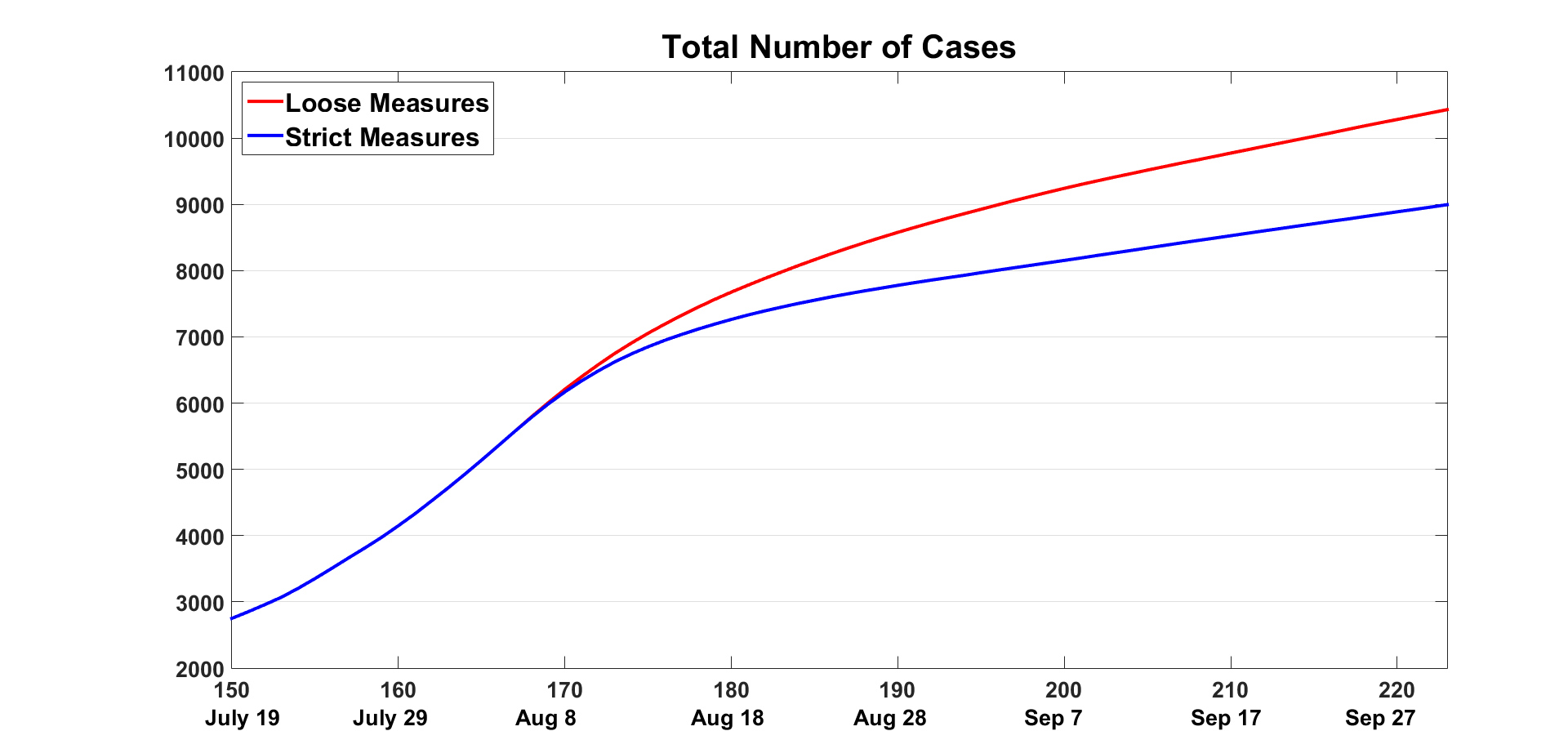}
\hspace*{0.2cm}
\includegraphics[width=7cm]{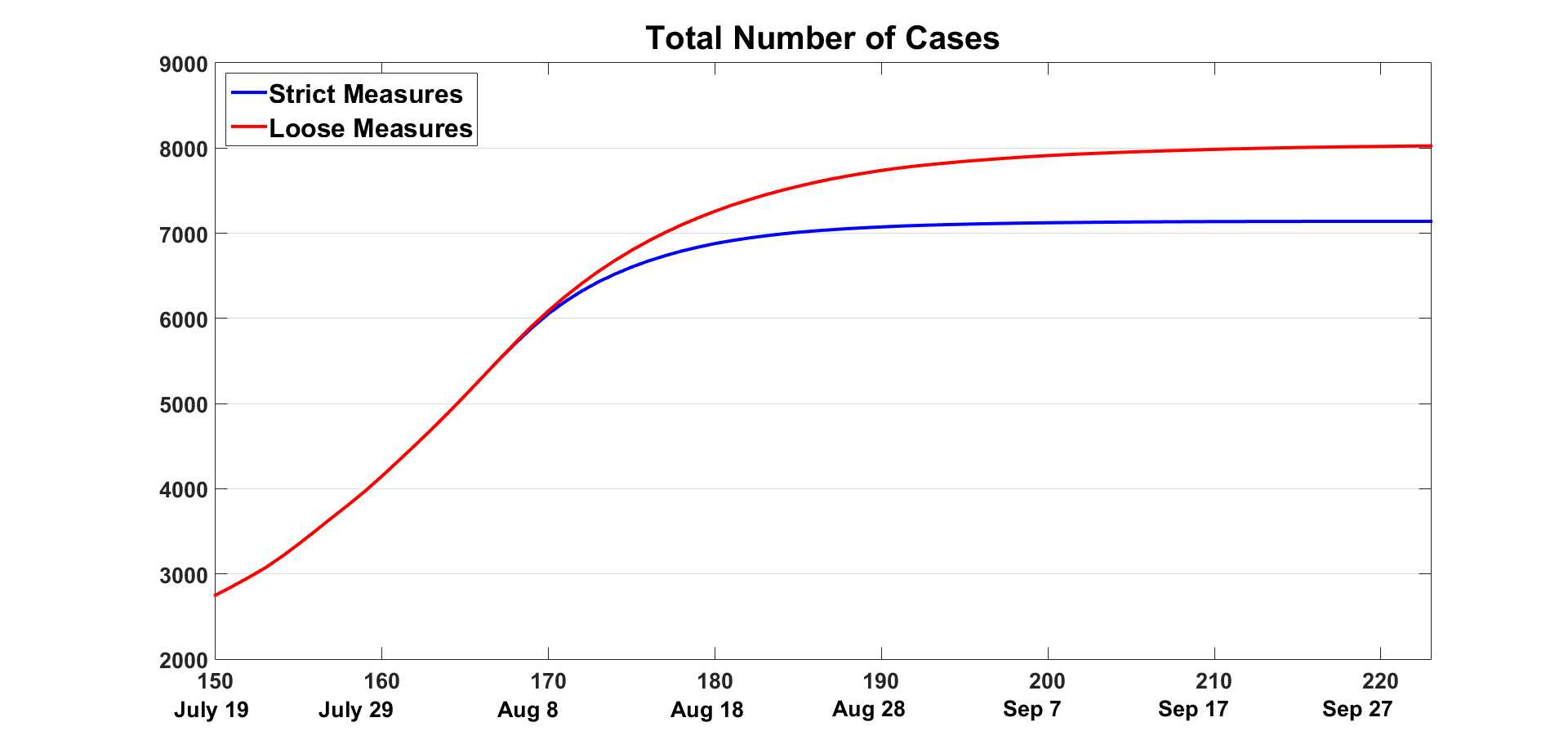}
\caption{Cumulative number of cases by September 30 under the four scenarios with open airport (left) and with closed airport (right).}\label{Cumul Scenarios}
\end{figure}

In Table \ref{tab:CumulAllScena}, we summarize the total number of cases for the four scenarios by the end of August.

\begin{minipage}{\linewidth}
\centering
\captionof{table}{Total number of cases on August 31 for the four scenarios.} \label{tab:CumulAllScena} 
\begin{tabular}{|l|c|c|c|c|}\toprule[1.5pt]
\textbf {Day} & \textbf {Scenario I} & \textbf {Scenario II} & \textbf {Scenario III} & \textbf {Scenario IV} \\\midrule
\hline
%{\bf Aug 1}& 4,724$\pm$1,358& 4,722$\pm$1,359& 4,702$\pm$1,359 & 4,702$\pm$1,359 \\
%\hline
%{\bf Aug 15}& 7,035$\pm$2,063& 7,323$\pm$2,175  & 6,736$\pm$2,037& 7,007$\pm$2,181\\
%\hline
{\bf Aug 31}& 7,893$\pm$2,193& 8,793$\pm$2,549 & 7,096$\pm$2,153& 7,808$\pm$2,530\\
\hline
\bottomrule[1.25pt]
\end {tabular}\par
\bigskip
%Should be a caption
\end{minipage}
\newline

%%%%%%%%%%%%%%%%%%%%%%%%%%%%%%%%%%%%%%%%%%%%%%%%%%%%%%%%%%%%%%%%%%%%%%%%%%%%%%%
\section{Conclusions and Discussion}\label{sec:conclusion}
We proposed, in the present work, a stochastic individual-based model for the spread of COVID-19 in Lebanon that takes into account some characteristics of the virus and the Lebanese population. On one hand, we considered probability distributions for the incubation period of the virus and for the available testing sensitivity. On the other hand, we took into account probability distributions for family size in the same household, new contacts met per day and the degree of  adherence of the individuals to mitigation measures. We simulated four different scenarios in which we compared the situation in two cases where the airport is closed during the last two weeks of August opposed to two cases where the airport is kept open and under different levels of strictness of mitigation measures. In all scenarios, the peak in the daily new cases is expected during the first week of August. If the airport is kept open, by the end of August an average of 70 new cases per day is expected under loose measures compared to  an average of 37 new cases per day under strict measures. However, if the airport is closed during the last two weeks of August, the average number of new cases per day is expected to drop to 20 under loose measures compared to an average of 6 cases under strict measures (see Figure \ref{Daily No Measures}). Consequently, a plateau is observed in the total number of cases by September (if the airport is closed) compared to a continuous increase in the number of cases during September if the airport is kept open (see Figure \ref{Cumul Scenarios}). \\
Despite the lack of census in Lebanon and the absence of public official reports containing real contact-tracing data the suggested model captured effectively the pattern of propagation of the virus in the country and is shedding light on the efficiency of mitigation measures imposed by policy makers.

%%*****************************************************
\bibliographystyle{plain}
\bibliography{biblioCorona1}
\end{document}